\documentclass[12pt,a4paper,dvips]{article}
\usepackage{a4p}
\usepackage{cite}
\usepackage{graphicx}
\usepackage{physics }
\usepackage{thisl3_titlePH,ifthen}
\usepackage{amsmath}
\usepackage{rotating}

\journalname{Phys. Lett. B}
\date{May 12, 2005}
\preprint{2005-021}

\newlength{\capindent}
\newlength{\capwidth}
\newlength{\figwidth}
\newcommand{\icaption}[2][!*!,!]{\hspace*{\capindent}%
  \begin{minipage}{\capwidth}
    \ifthenelse{\equal{#1}{!*!,!}}%
      {\caption{#2}}%
      {\caption[#1]{#2}}
  \end{minipage}}
\renewcommand{\deg}{\ensuremath{^\mathrm{o}}}

\newcommand{\peneg}{\ensuremath{{\mathrm{e}^{-}}}}
\newcommand{\pepos}{\ensuremath{{\mathrm{e}^{+}}}}

\newcommand{\lum}{\ensuremath{\mathcal{L}}}

\begin{document}
\begin{titlepage}
\title{Measurement of  the Running of
the Electromagnetic Coupling at Large Momentum-Transfer at LEP}
\author{The L3 Collaboration}
%
%

\begin{abstract}
 
The evolution of the electromagnetic coupling, $\alpha$, in the
momentum-transfer range $1800\GeV^2 < -Q^2 < 21600\GeV^2$ is studied
with about 40\,000 Bhabha-scattering events collected with the L3 detector
at LEP at centre-of-mass energies $\sqrt{s}= 189-209 \GeV$. The
running of $\alpha$ is parametrised as:
\begin{displaymath}
  \alpha(Q^2) = \frac{\alpha_0}{1-C\Delta\alpha(Q^2)},
\end{displaymath}
where $\alpha_0\equiv\alpha(Q^2=0)$ is the fine-structure constant and
$C=1$ corresponds to the evolution expected in QED. A fit to the
differential cross section of the $\ee\ra\ee$ process for scattering
angles in the range $|\cos\theta|<0.9$ excludes the hypothesis of
a constant value of $\alpha$, $C=0$, and validates the QED prediction
with the result:
\begin{displaymath}
C  =  1.05 \pm 0.07 \pm 0.14,
\end{displaymath}
where the first uncertainty is statistical and the second systematic.
\end{abstract}
%
%
\submitted
\end{titlepage}

%
\section{Introduction}
%

\begin{figure}[p]
  \begin{center}
    \includegraphics[width=\textwidth]{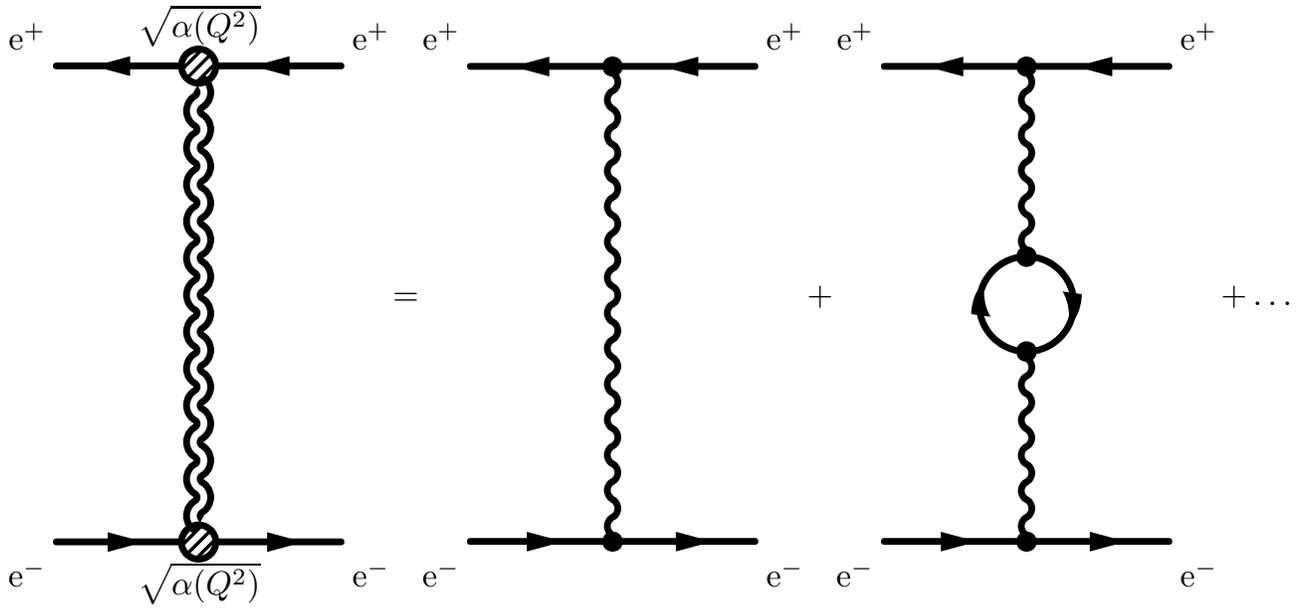}\\
  \end{center}
  \caption{$t$-channel Feynman diagrams contributing to Bhabha
           scattering. Diagrams with virtual-fermion
           vacuum-polarisation insertions generate an electromagnetic
           coupling $\alpha(Q^2)$. The sum of all diagrams including
           zero, one, two or more vacuum-polarisation insertions is
           denoted by the diagram to the left with the double-wavy
           photon propagator.}
  \label{fig:1}
\end{figure}

A fundamental consequence of quantum field theory is that the value of
the electromagnetic coupling, $\alpha$, depends on, or {\it runs}
with, the squared momentum transfer, $Q^2$. This phenomenon is due to
higher momentum-transfers probing virtual-loop corrections to the
photon propagator. This process of {\it vacuum polarisation} is
sketched in Figure~\ref{fig:1}.  In QED, the dependence of $\alpha$ on
$Q^2$ is described as~\cite{ref:running}:
\begin{equation}
  \alpha(Q^{2}) = \frac{\alpha_0}{1-\Delta\alpha(Q^{2})},
  \label{eq:aqsq}
\end{equation}
where the fine-structure constant, $\alpha_0\equiv\alpha(Q^2=0)$, is a
fundamental quantity of Physics. It is measured with high accuracy in
solid-state processes and via the study of the anomalous magnetic
moment of the electron to be $1/\alpha_0 =
137.03599911\pm0.00000046$~\cite{ref:codata}.  The contributions to
$\Delta\alpha(Q^{2})$ from lepton loops are precisely
predicted~\cite{ref:steinhauser}, while those from quark loops are
difficult to calculate due to non-perturbative QCD effects. They are
estimated using dispersion-integral techniques~\cite{ref:gatto} and
information from the $\ee \ra {hadrons}$ cross section. At the scale
of the Z-boson mass, recent calculations yield $\alpha^{-1}(\MZ^{2}) =
128.936 \pm 0.046$~\cite{ref:burkhardt_new}. Similar results, with
smaller uncertainty, are found by other evaluations using stronger
theoretical assumptions.  For example, Reference~\citen{ref:troconiz}
obtains $\alpha^{-1}(\MZ^{2}) = 128.962 \pm 0.016$.

The running of $\alpha$ was studied at $\epem$ colliders both in the
{\it time-like} region, $Q^2>0$, and the {\it space-like} region,
$Q^2<0$. The first measurement in the time-like region was performed
by the TOPAZ Collaboration at TRISTAN for $Q^2=3338\GeV^2$ by
comparing the cross sections of the $\epem\ra\epem$ and
$\epem\ra\epem\mu^+\mu^-$ processes~\cite{ref:topaz}. The OPAL
Collaboration at LEP exploited the different sensitivity to
$\alpha(Q^2)$ of the cross sections of the $\epem\ra\mu^+\mu^-$,
$\epem\ra\tau^+\tau^-$ and $\epem\ra\rm q \bar{q}$ processes above the
Z resonance to determine
$\alpha(37236\GeV^2)$~\cite{ref:opal1}. Information on
$\alpha(\MZ^{2})$ is also extracted from the couplings of the Z boson
to fermion pairs~\cite{sunandaew}.

Bhabha scattering at $\epem$ colliders, $\epem\ra\epem$, gives access
to the running of $\alpha$ in the space-like region. In addition, like
other processes dominated by $t$-channel photon exchange, it has little
dependence on weak corrections.  The
four-momentum transfer in Bhabha scattering depends on $s$ and on the
scattering angle, $\theta$: $Q^{2} = t \simeq -s(1-\cos\theta)/2 < 0$.
Small-angle and large-angle Bhabha scattering allow to probe the
running of $\alpha$ in different $Q^2$ ranges.

LEP detectors were equipped with luminosity monitors, high-precision
calorimeters located close to the beam pipe and designed to measure
small-angle Bhabha scattering in order to determine the integrated
luminosity collected by the experiments.  The L3 collaboration first
established the running of $\alpha$ in the range
$2.10\GeV^2<-Q^2<6.25\GeV^2$~\cite{l3-197} by comparing event counts in
different regions of its luminosity monitor. More recently, the OPAL
Collaboration studied the similar range
$1.81\GeV^2<-Q^2<6.07\GeV^2$~\cite{ref:opal2}.

The running of $\alpha$ in large-angle Bhabha scattering was first
investigated by the VENUS Collaboration at TRISTAN in the range
$100\GeV^2<-Q^2<2916\GeV^2$~\cite{ref:venus}. Later, the L3
Collaboration studied the same process at
$\sqrt{s}=189\GeV$ for scattering angles $0.81 < |\cos\theta| < 0.94$,
probing the range $12.25\GeV^2<-Q^2<3434\GeV^2$~\cite{l3-197}.

This Letter investigates the running of $\alpha$ by studying the
differential cross section for Bhabha scattering at LEP at $\rts =
189-209 \GeV$ for scattering angles such that $|\cos\theta| < 0.9$. Less than
1\% of the events scatter backwards, $\cos\theta < 0$, and
this analysis effectively probes the region $ 1800 \GeV^{2} < -Q^{2} <
21600 \GeV^{2}$, extending and complementing previous space-like
studies.

%
\section{Analysis Strategy}
%

In the following, the running of $\alpha$ is described by a free
parameter, $C$, defined according to:
\begin{equation}
  \alpha(Q^2) = \frac{\alpha_0}{1-C\Delta\alpha(Q^2)},
  \label{eq:cdef}
\end{equation}
where the parametrisation of Reference~\citen{ref:burkhardt_new} is
used for the term $\Delta\alpha(Q^2)$. A value of $C$ consistent with
$C=1$ would indicate that data follow the behaviour predicted by QED,
while the hypothesis $\alpha=\alpha_0$, with no dependence on
$Q^2$, corresponds to $C=0$.

The value of $C$ is derived by a study of the measured differential
cross section of the $\epem\ra\epem$ process, ${\rm d}\sigma / {\rm
d}\cos\theta$. This quantity depends on $C$ through the measured
integrated luminosity, ${\cal L}(C)$, which is calculated from the
expected cross section of the $\epem\ra\epem$ process for small
scattering angles. The measurements used in the following are obtained
under the Standard Model hypothesis, $C=1$, as:
\begin{eqnarray}
  \frac{{\rm d} \sigma(1)}{{\rm d}\cos\theta} & = & \frac{
  N(\cos\theta)}{\Delta\cos\theta}
  \frac{1}{\lum(1)\varepsilon(\cos\theta)},
\end{eqnarray}
where $N(\cos\theta)$ is the number of events observed in a given
$\cos\theta$ range, of width $\Delta\cos\theta$, with average
acceptance $\varepsilon(\cos\theta)$.  The
measured integrated luminosity depends on $C$ as:
\begin{eqnarray}
  \lum(C) \equiv \frac{N_{L}}{\sigma_{L}(C)\varepsilon_{L}(C)} & = &
          \lum(1)\frac{\sigma_{L}(1)\varepsilon_{L}(1)}
          {\sigma_{L}(C)\varepsilon_{L}(C)},
\end{eqnarray}
where $N_L$ is the number of events observed in the fiducial volume of
the luminosity monitor, $\sigma_{L}(C)$ is the corresponding
$\epem\ra\epem$ cross section for a given value of $C$ and
$\varepsilon_{L}(C)$ is the detector acceptance. This acceptance may
depend on $C$ due to the combined effect of small angular
anisotropies of detector efficiencies and the dependence of the
predicted differential cross section on $C$.  These changes in the
acceptance are found to have negligible impact on the results
presented below.

The value of the parameter $C$ is extracted by comparing the measured differential
cross section to the
theoretical prediction as a function of $C$, ${\rm d}\sigma^{th}(C) /
{\rm d}\cos\theta$, derived as:
\begin{eqnarray}
  \frac{{\rm d} \sigma^{th}(C)}{{\rm d}\cos\theta} & \equiv &
          \frac{{\rm d} \sigma^{th}(1)}{{\rm d}\cos\theta}
          \frac{\lum(1)}{\lum(C)},
\end{eqnarray}
where ${\rm d}\sigma^{th}(1) / {\rm d}\cos\theta$ is the Standard
Model prediction, discussed in Reference~\citen{ref:yellow}. The value
of $\lum(1)$ is derived by using the BHLUMI Monte Carlo program~\cite{BHLUMI}.
The dependence of ${\rm d}\sigma^{th}(C) / {\rm d}\cos\theta$ and
${\cal L}(C)$ on $C$ is implemented by means of the BHWIDE
Monte Carlo program~\cite{BHWIDE}. The differential cross section is
factorised as:
\begin{equation}
  \frac{{\rm d}\sigma^{th}(C)}{{\rm d}\cos\theta} \equiv \frac{{\rm
  d}\sigma^{Born}(C)}{{\rm d}\cos\theta}
  F_{rad}\left(\cos\theta\right),
\end{equation}
where ${\rm d}\sigma^{Born}(C) / {\rm d}\cos\theta$ is the tree-level
differential cross section, which has a simple analytical form. The
term $F_{rad}\left(\cos\theta\right)$ parametrises initial-state and
final-state radiation effects, dominated by real-photon emission, as
implemented in BHWIDE. It is verified that
$F_{rad}\left(\cos\theta\right)$ has a negligible dependence on the
spread of $\sqrt{s}$ considered in this analysis and, most important,
on $C$.

%
\section{Cross Section Measurement}
%

The data were collected at LEP by the L3 detector~\cite{l3,egap} in
the years from 1998 through 2000. They correspond to an integrated
luminosity of 607.4~\pb\ and are grouped in eight intervals of $\rts$
with the average values and corresponding integrated luminosities listed
in Table~\ref{tab:1}.

Events from the $\epem\ra\epem$ process are selected as described in
Reference~\citen{ref:fpp_paper}. Electrons and positrons are
identified as clusters in the BGO electromagnetic calorimeter, matched
with tracks in the central tracker. In the barrel region of the
detector, $|\cos\theta|<0.72$, the energy of the most energetic
cluster must satisfy $E_1> 0.25\sqrt{s}$, while the energy of the other
cluster must satisfy $E_2>20\GeV$. In the endcap region,
$0.81<|\cos\theta|<0.98$, these criteria are relaxed to
$E_1>0.2\sqrt{s}$ and $E_2>10\GeV$. Events with clusters in the
transition region between the barrel and endcap regions,
$0.72<|\cos\theta|<0.81$, instrumented with a lead and
scintillating-fiber calorimeter~\cite{egap}, are rejected.  To
suppress contributions from events with high-energy initial-state
radiation, the complement to $180^\circ$ of the angle between the two
clusters, the acollinearity, $\zeta$, is required to be less than
$25\deg$. The number of events observed at different values of
$\sqrt{s}$ is shown in Table~\ref{tab:1} together with the Monte Carlo
expectations for signal and background.

The $\ee\ra\ee$ process is simulated with the BHWIDE Monte Carlo
generator assuming $C=1$. Background processes are described with the
following Monte Carlo generators: KORALZ~\cite{KORALZ} for
$\epem\ra\tau^+\tau^-$, KORALW~\cite{KORALW} for $\epem\ra\rm W^+W^-$,
PYTHIA~\cite{PYTHIA} for $\epem\ra\rm Z\epem$, DIAG36~\cite{DIAG36}
for $\epem\ra\epem\epem$, GGG~\cite{GGG} for
$\epem\ra\gamma\gamma\gamma$ and TEEGG~\cite{TEEGG} for
$\epem\ra\epem\gamma$ events where one fermion is scattered into the
beam pipe and the photon is in the detector.  The L3 detector response
is simulated using the GEANT package~\cite{geant}, which describes
effects of energy loss, multiple scattering and showering in the
detector.  Time-dependent detector inefficiencies, as monitored during
the data-taking period, are included in the simulation.

Systematic effects, such as charge confusion, are
reduced by folding the differential cross section into ${\rm d} \sigma/
{\rm d}|\cos\theta|$, which is defined as:
\begin{equation}
  \frac{{\rm d}\sigma}{{\rm d}|\cos\theta|} \equiv \frac{{\rm
  d}\sigma}{{\rm d}\cos\theta}\big|_{\cos\theta<0} +\frac{{\rm
  d}\sigma}{{\rm d}\cos\theta}\big|_{\cos\theta>0}.
\end{equation}
This differential cross section is measured in the fiducial
volume defined by:
\begin{gather}
        12\deg < \theta_{\peneg,\pepos} < 168\deg\\ |\cos\theta| < 0.9
        \\ \zeta < 25\deg
\end{gather}
where $\theta_{\peneg}$ and $\theta_{\pepos}$ are the polar angles of
the electron and the positron, respectively. The value of $\cos\theta$
is derived as:
\begin{equation}
  \cos\theta \equiv
  \frac{\sin\left|\theta_\pepos-\theta_\peneg\right|}{\sin\theta_\peneg+\sin\theta_\pepos}.
\end{equation}

\begin{figure}[p]
  \begin{center}
    \includegraphics[width=1.1\textwidth]{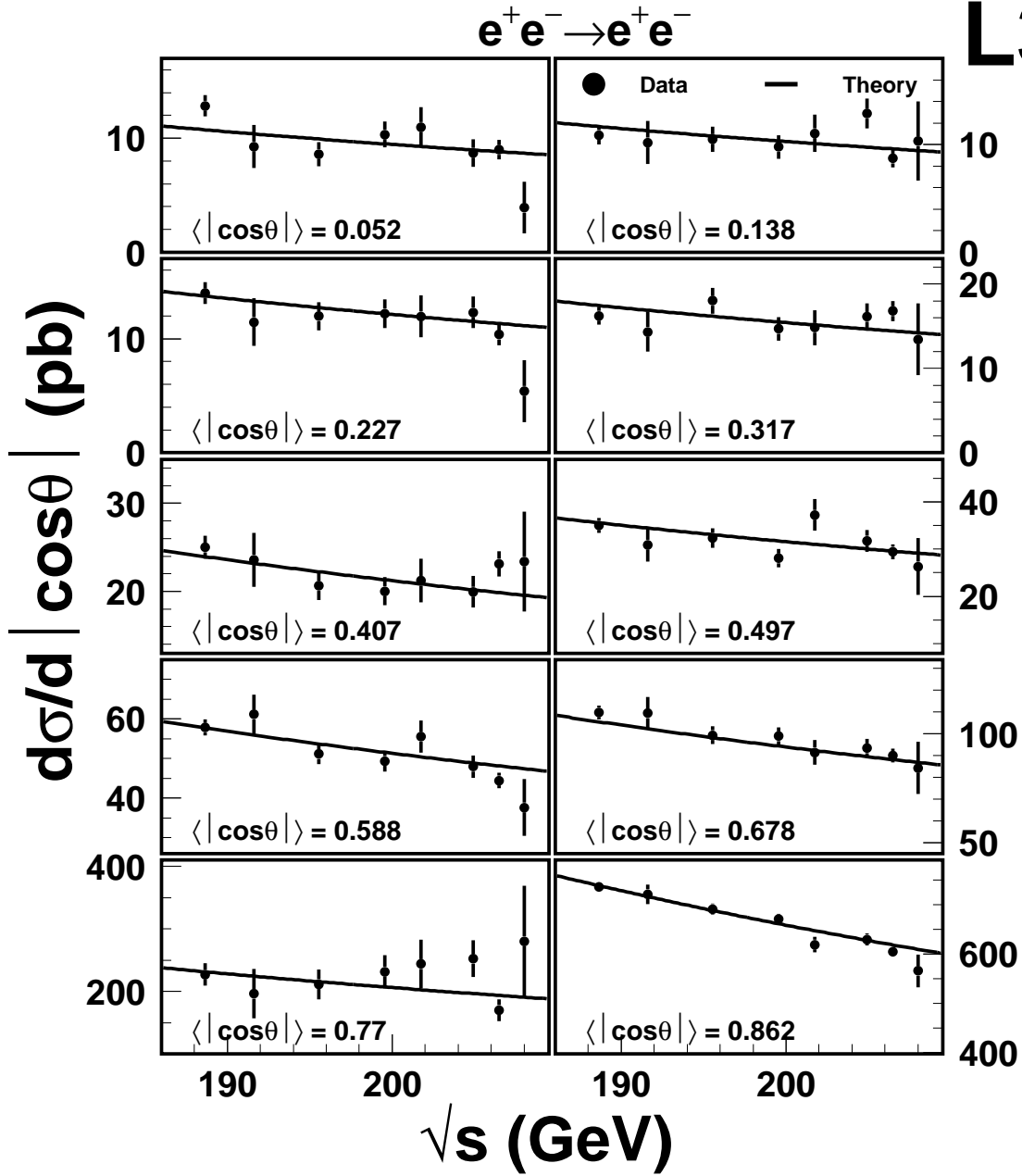}
  \end{center}
  \caption{Measured Bhabha differential cross-sections 
           for the ten  $|\cos\theta|$ intervals used 
           in the study as a function of the centre-of-mass energy
           $\rts$. The Standard Model predictions are represented by
           the solid lines.}
  \label{fig:2}
\end{figure}

Ten intervals of $|\cos\theta|$ are considered for each of the eight
values of $\sqrt{s}$, for a total of 80 independent measurements.
Table~\ref{tab:2} and Figure~\ref{fig:2} present the measurements of
${\rm d} \sigma/ {\rm d}|\cos\theta|$ and the Standard Model
expectations. The larger uncertainties in the interval $0.72-0.81$ are
due to the transition region between the barrel and the endcap regions.

%
\section{Results}
%
 
\begin{figure}[p]
  \begin{center}
    \includegraphics[width=\textwidth]{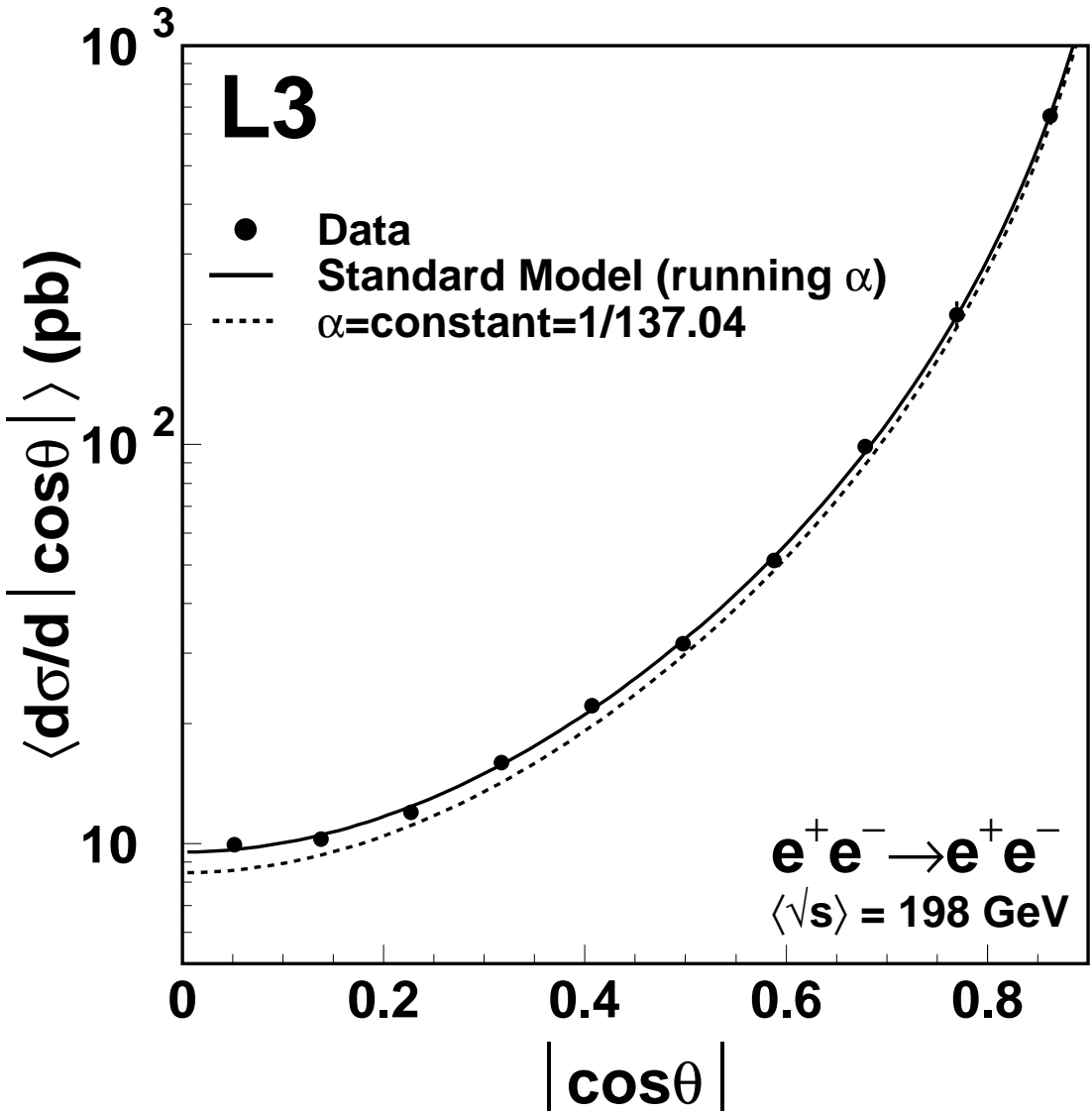}
  \end{center}
  \caption{Measured differential cross-section as a function of
  $|\cos\theta|$. Data at different centre-of-mass energies are
  combined at the luminosity-averaged centre-of-mass energy
  $\langle\rts\rangle=198 \GeV$. The predictions in case of a running
  electromagnetic coupling and for a constant value $\alpha=\alpha_0$
  are also shown.}
  \label{fig:3}
\end{figure}

\begin{figure}[p]
  \begin{center}
    \includegraphics[width=\textwidth]{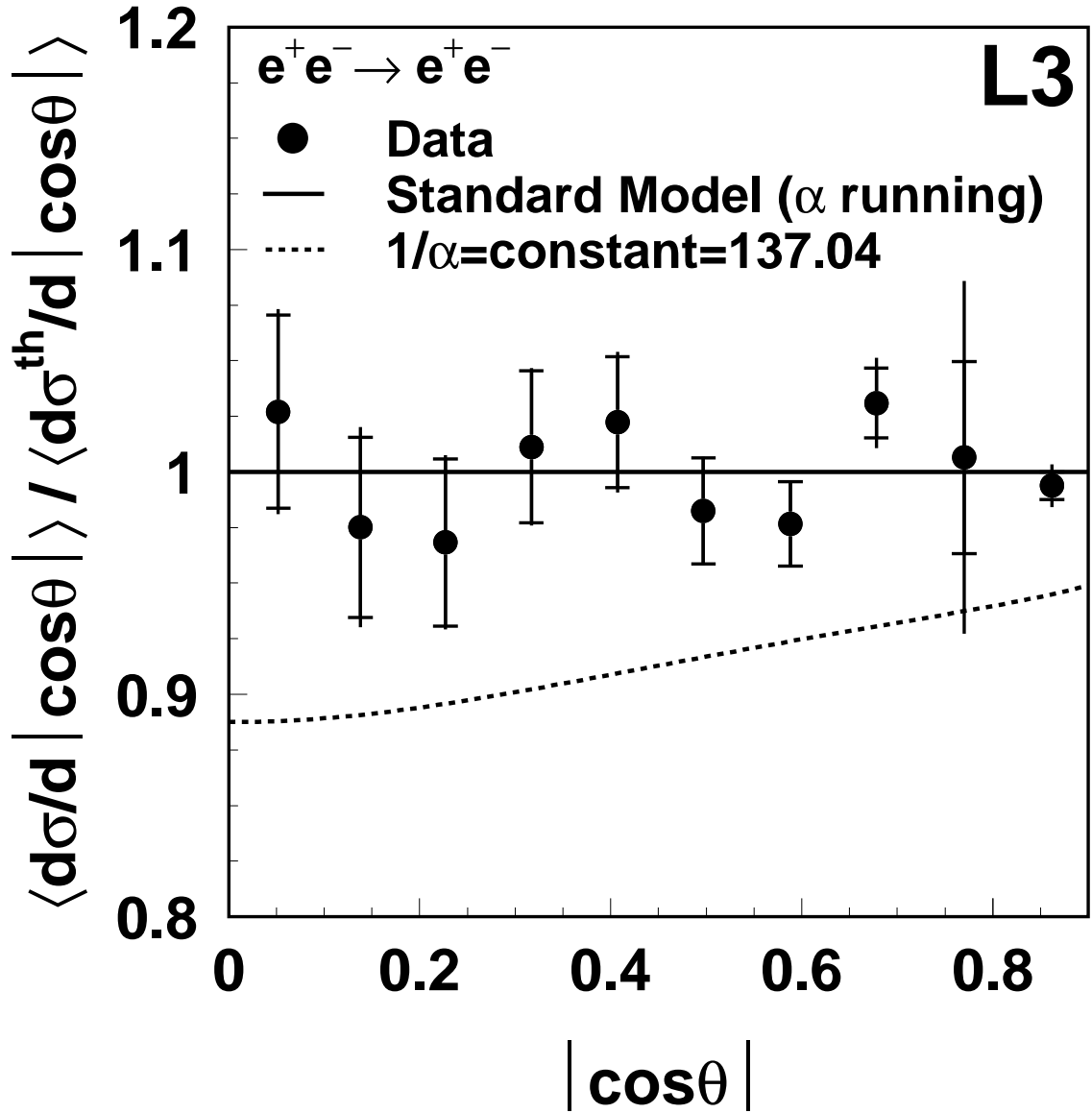}
  \end{center}
  \caption{Ratio between the measured Bhabha differential cross
  section as a function of $|\cos\theta|$ and the Standard Model
  expectations including a running electromagnetic coupling. Data at
  different centre-of-mass energies are combined at the
  luminosity-averaged centre-of-mass energy $\langle\rts\rangle=198
  \GeV$. The inner error bars represent statistical uncertainties and
  the full error bars the sum in quadrature of the statistical and
  systematic uncertainties. The predictions for a constant value of
  $\alpha=\alpha_0$ are also shown.  }
  \label{fig:4}
\end{figure}

Figures~\ref{fig:3} and~\ref{fig:4} compare the combined differential
cross section at the average centre-of-mass energy
$\langle\sqrt{s}\rangle=198\GeV$ with the Standard Model prediction,
corresponding to $C=1$, and with a constant value of $\alpha$,
corresponding to $C=0$. The data favour the hypothesis $C=1$ over the
hypothesis $C=0$, as also presented in Table~\ref{tab:3}.

The value of $C$ is extracted by comparing the 80 measurements of
${\rm d} \sigma/ {\rm d}|\cos\theta|$ with the theoretical
expectations ${\rm d}\sigma^{th}(C) / {\rm d}\cos\theta$ in a $\chi^2$
fit with the result:
\begin{displaymath}
C = 1.06 \pm 0.07,
\end{displaymath}
where the quoted uncertainty is statistical only.  Several sources of
systematic uncertainties are then considered.
\begin{itemize}
\item The theoretical expectations for ${\rm d}\sigma^{th}(1) / {\rm
  d}\cos\theta$ have an uncertainty which varies from 0.5\% in the
  endcap region to 1.5\% in the barrel region~\cite{ref:yellow,BHWIDE}.
\item The measurements of ${\rm d} \sigma/ {\rm d}|\cos\theta|$ are
  affected by a systematic uncertainty, dominated by the
  event-selection procedure, which varies between 1\% and 10\%, as
  listed in Table~\ref{tab:2}~\cite{ref:fpp_paper}.
\item An uncertainty between 0.2\% and 1.5\% is assigned to
  $F_{rad}\left(\cos\theta\right)$, as a function of $\cos\theta$, in
  order to account for possible higher-order effects not included in
  the BHWIDE parametrisation.
\item Migration effects among the different $\cos\theta$ bins are
  found to be negligible due to the large bin size and the good detector
  resolution.
\end{itemize}

Systematic uncertainties are conservatively treated as fully
correlated and the fit is repeated including both statistical and
systematic uncertainties with the result:
\begin{displaymath}
C = 1.05 \pm 0.07 \pm 0.14,
\end{displaymath}
where the first uncertainty is statistical and the second
systematic. A breakdown of the systematic uncertainty is presented in
Table~\ref{tab:4}. This result is in agreement with the Standard Model
expectation, $C=1$. The quality of the fit is satisfactory, with a
$\chi^2$ of 91.9 for 79 degrees of freedom, corresponding to a
confidence level of 17\%.  The hypothesis of a value of $\alpha$ which
does not depend on $Q^2$, $C=0$, is totally excluded with a $\chi^2$ of
316 for 80 degrees of freedom, corresponding to a a confidence level
of $10^{-29}$.

%
\section{Discussion}
%
 
The result presented above establishes the evolution of the
electromagnetic coupling with $-Q^2$ in the range $ 1800 \GeV^{2} < -Q^{2} <
21600 \GeV^{2}$. This finding extends and complements studies based on small-angle
Bhabha scattering by the L3~\cite{l3-197} and OPAL~\cite{ref:opal2}
Collaborations, which studied the regions $2.10\GeV^2<-Q^2<6.25\GeV^2$
and $1.81\GeV^2<-Q^2<6.07\GeV^2$, respectively. The advantage of
large-angle Bhabha scattering, investigated in this Letter, is to
probe large values of $-Q^2$, while studies of small-angle Bhabha
scattering at lower values of $-Q^2$ benefit from a larger cross section
and thus statistical accuracy.  The experimental systematic
uncertainties of  measurements in the two $-Q^2$ regions are
implicitly different. At large $-Q^2$, they are dominated by the
selection of Bhabha events in the large-angle calorimeters, while at
low $-Q^2$ they mostly arise from the event reconstruction in the
luminosity monitors and from effects of the material traversed by
electrons and positrons before their detection. Both studies, at large
and low  $-Q^2$, are affected by theoretical uncertainties on
the differential cross section of Bhabha scattering, although in
different angular regions

\begin{figure}[p]
  \begin{center}
    \includegraphics[width=\textwidth]{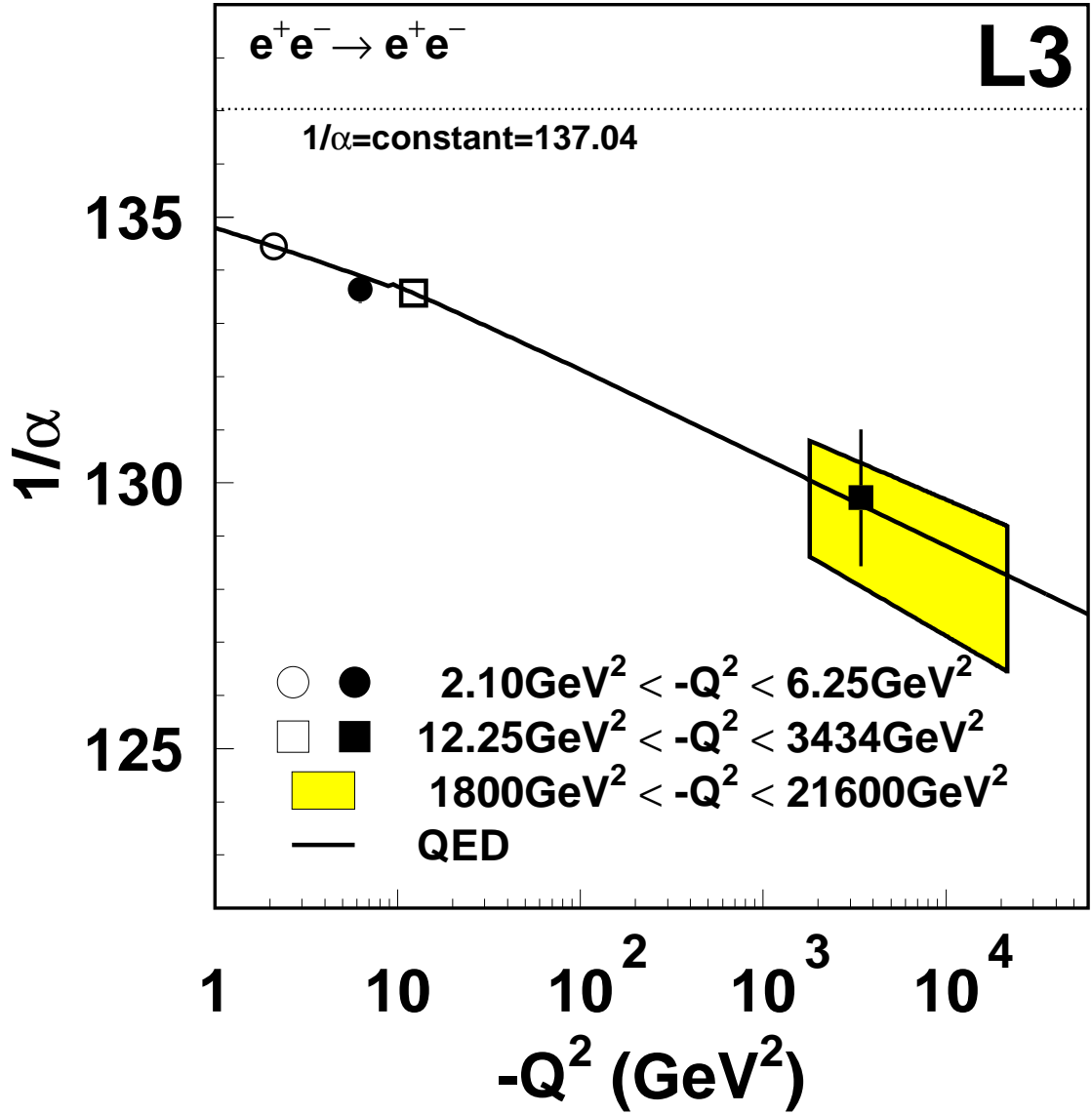}\\
  \end{center}
  \caption{Evolution of the electromagnetic coupling with $Q^2$
determined from the present measurement of $C$ for $ 1800 \GeV^{2} <
-Q^{2} < 21600 \GeV^{2}$, yellow band, and from previous data for
Bhabha scattering at $2.10\GeV^2<-Q^2<6.25\GeV^2$ and
$12.25\GeV^2<-Q^2<3434\GeV^2$~\protect\cite{l3-197}. The open symbols
indicate the values of $Q^2$ where $\alpha(Q^2)$ was fixed to the QED
predictions\protect\cite{ref:burkhardt_new} in order to infer the
values of $\alpha(Q^2)$ shown by the full symbols. These QED
predictions are shown by the solid line.}
  \label{fig:5}
\end{figure}

\begin{figure}[p]
  \begin{center}
    \includegraphics[width=\textwidth]{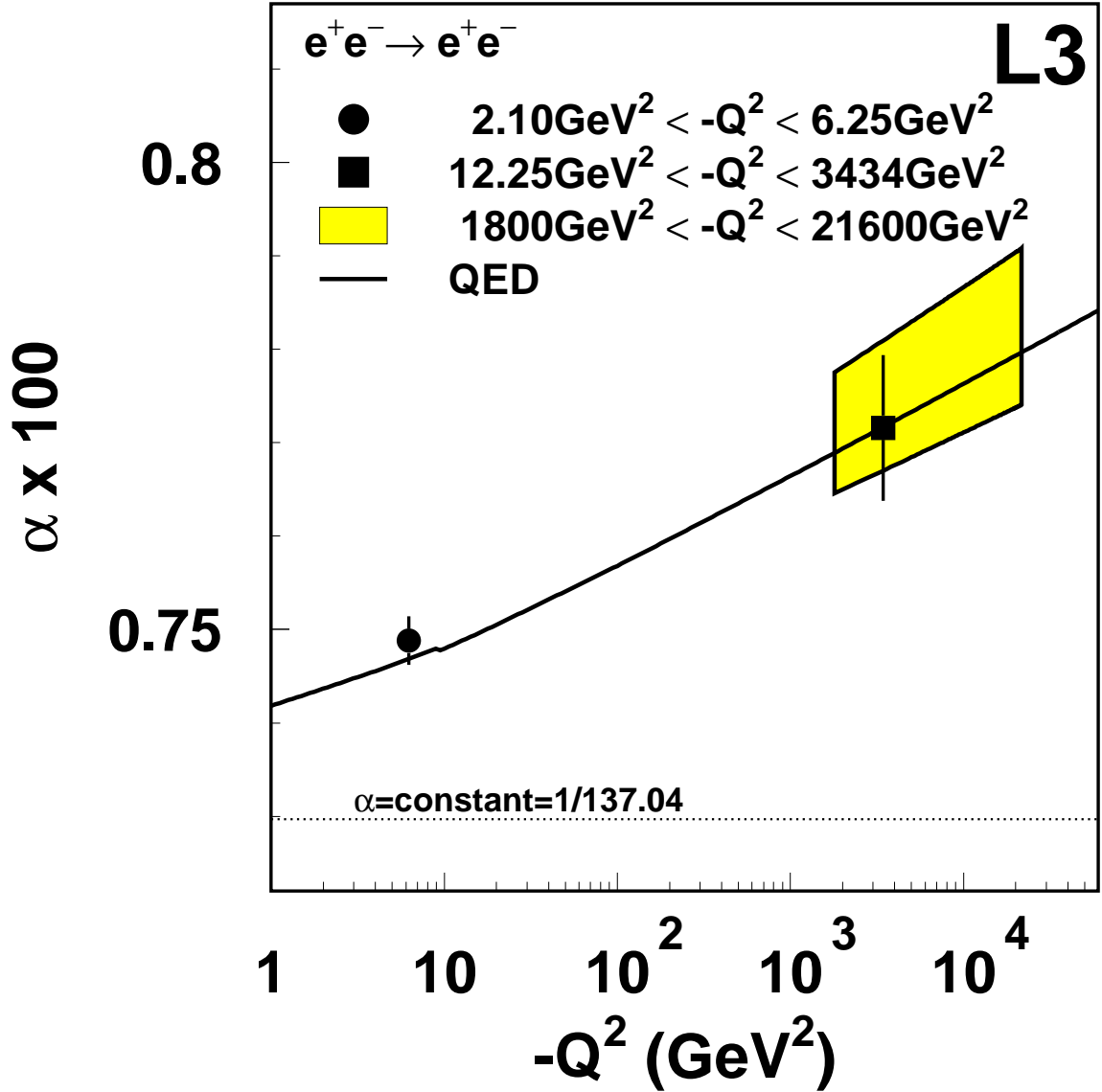}\\
  \end{center}
  \caption{Evolution of the electromagnetic coupling with $Q^2$
determined from the present measurement of $C$ for $ 1800 \GeV^{2} <
-Q^{2} < 21600 \GeV^{2}$, yellow band, and from previous data for
Bhabha scattering at $2.10\GeV^2<-Q^2<6.25\GeV^2$ and
$12.25\GeV^2<-Q^2<3434\GeV^2$~\protect\cite{l3-197}, full symbols. The
solid line represent the QED
predictions\protect\cite{ref:burkhardt_new}.}
  \label{fig:6}
\end{figure}

Figures~\ref{fig:5}~and~\ref{fig:6} present the evolution of the
electromagnetic coupling with $-Q^2$. A band for $1800 \GeV^{2} <
-Q^{2} < 21600 \GeV^{2}$ shows the 68\% confidence level result from
this analysis. It is derived by inserting the measured value of $C$
with its errors in Equation~(\ref{eq:cdef}) together with the QED
predictions of Reference~\citen{ref:burkhardt_new}. The results from
previous L3 data for Bhabha scattering at $2.10\GeV^2<-Q^2<6.25\GeV^2$
and $12.25\GeV^2<-Q^2<3434\GeV^2$~\cite{l3-197} are also shown. These
two measurements are not absolute measurements of the electromagnetic
coupling but differences between the values of $\alpha(Q^2)$ at the
extreme of the $Q^2$ ranges~\cite{l3-197}:
\begin{eqnarray}
    \alpha^{-1}(-2.10 \GeV^{2}) - \alpha^{-1}(-6.25 \GeV^{2}) & = &
    0.78 \pm 0.26  
    \label{eq:old1} \\
    \alpha^{-1}(-12.25 \GeV^{2}) - \alpha^{-1}(-3434
    \GeV^{2}) & = & 3.80 \pm 1.29.
    \label{eq:old2}
\end{eqnarray}
The results in Figure~\ref{fig:5} are obtained by fixing the
values of $\alpha(-2.10\GeV^2 )$ and $\alpha(-12.25\GeV^2 )$ to the
QED predictions of Reference~\citen{ref:burkhardt_new} in order to
extract the values of $\alpha(-6.25\GeV^2 )$ and $\alpha(-3434\GeV^2)$
from Equations~(\ref{eq:old1}) and~(\ref{eq:old2}).  The results shown
in Figure~\ref{fig:6} are obtained by first determining the values of
$\alpha(-2.10\GeV^2 )$ and $\alpha(-12.25\GeV^2 )$ from the measured
value of $C$ and from Equation~(\ref{eq:cdef}) and then extracting the
values of $\alpha(-6.25\GeV^2 )$ and $\alpha(-3434\GeV^2)$ from
Equations~(\ref{eq:old1}) and~(\ref{eq:old2}). This procedure relies
on the assumption that the measured value of $C$ also describes the
running of the electromagnetic coupling for lower values of $-Q^2$.
Both figures provide an impressive evidence of the running of the
electromagnetic coupling in the energy range accessible at LEP.

%
%

%
%
\newpage
\typeout{   }     
\typeout{Using author list for paper 287 -  }
\typeout{$Modified: Jul 15 2001 by smele $}
\typeout{!!!!  This should only be used with document option a4p!!!!}
\typeout{   }
%
%
%
%
%
%

\newcount\tutecount  \tutecount=0
\def\tutenum#1{\global\advance\tutecount by 1 \xdef#1{\the\tutecount}}
\def\tute#1{$^{#1}$}
\tutenum\aachen            
\tutenum\nikhef            
\tutenum\mich              
\tutenum\lapp              
\tutenum\basel             
\tutenum\lsu               
\tutenum\beijing           
\tutenum\bologna           
\tutenum\tata              
\tutenum\ne                
\tutenum\bucharest         
\tutenum\budapest          
\tutenum\mit               
\tutenum\panjab            
\tutenum\debrecen          
\tutenum\dublin            
\tutenum\florence          
\tutenum\cern              
\tutenum\wl                
\tutenum\geneva            
\tutenum\hamburg           
\tutenum\hefei             
\tutenum\lausanne          
\tutenum\lyon              
\tutenum\madrid            
\tutenum\florida           
\tutenum\milan             
\tutenum\moscow            
\tutenum\naples            
\tutenum\cyprus            
\tutenum\nymegen           
\tutenum\caltech           
\tutenum\perugia           
\tutenum\peters            
\tutenum\cmu               
\tutenum\potenza           
\tutenum\prince            
\tutenum\riverside         
\tutenum\rome              
\tutenum\salerno           
\tutenum\ucsd              
\tutenum\sofia             
\tutenum\korea             
\tutenum\taiwan            
\tutenum\tsinghua          
\tutenum\purdue            
\tutenum\psinst            
\tutenum\zeuthen           
\tutenum\eth               

{
\parskip=0pt
\noindent
{\bf The L3 Collaboration:}
\ifx\selectfont\undefined
 \baselineskip=10.8pt
 \baselineskip\baselinestretch\baselineskip
 \normalbaselineskip\baselineskip
 \ixpt
\else
 \fontsize{9}{10.8pt}\selectfont
\fi
\medskip
\tolerance=10000
\hbadness=5000
\raggedright
\hsize=162truemm\hoffset=0mm
\def\r{\rlap,}
\noindent

P.Achard\r\tute\geneva\ 
O.Adriani\r\tute{\florence}\ 
M.Aguilar-Benitez\r\tute\madrid\ 
J.Alcaraz\r\tute{\madrid}\ 
G.Alemanni\r\tute\lausanne\
J.Allaby\r\tute\cern\
A.Aloisio\r\tute\naples\ 
M.G.Alviggi\r\tute\naples\
H.Anderhub\r\tute\eth\ 
V.P.Andreev\r\tute{\lsu,\peters}\
% ...
\newpage
\begin{list}{A}{\itemsep=0pt plus 0pt minus 0pt\parsep=0pt plus 0pt minus 0pt
                \topsep=0pt plus 0pt minus 0pt}
\item[\aachen]
 III. Physikalisches Institut, RWTH, D-52056 Aachen, Germany$^{\S}$
\item[\nikhef] National Institute for High Energy Physics, NIKHEF, 
     and University of Amsterdam, NL-1009 DB Amsterdam, The Netherlands
\item[\mich] University of Michigan, Ann Arbor, MI 48109, USA
\item[\lapp] Laboratoire d'Annecy-le-Vieux de Physique des Particules, 
     LAPP,IN2P3-CNRS, BP 110, F-74941 Annecy-le-Vieux CEDEX, France
\item[\basel] Institute of Physics, University of Basel, CH-4056 Basel,
     Switzerland
\item[\lsu] Louisiana State University, Baton Rouge, LA 70803, USA
\item[\beijing] Institute of High Energy Physics, IHEP, 
  100039 Beijing, China$^{\triangle}$ 
\item[\bologna] University of Bologna and INFN-Sezione di Bologna, 
     I-40126 Bologna, Italy
\item[\tata] Tata Institute of Fundamental Research, Mumbai (Bombay) 400 005, India
\item[\ne] Northeastern University, Boston, MA 02115, USA
\item[\bucharest] Institute of Atomic Physics and University of Bucharest,
     R-76900 Bucharest, Romania
\item[\budapest] Central Research Institute for Physics of the 
     Hungarian Academy of Sciences, H-1525 Budapest 114, Hungary$^{\ddag}$
\item[\mit] Massachusetts Institute of Technology, Cambridge, MA 02139, USA
\item[\panjab] Panjab University, Chandigarh 160 014, India
\item[\debrecen] KLTE-ATOMKI, H-4010 Debrecen, Hungary$^\P$
\item[\dublin] Department of Experimental Physics,
  University College Dublin, Belfield, Dublin 4, Ireland
\item[\florence] INFN Sezione di Firenze and University of Florence, 
     I-50125 Florence, Italy
\item[\cern] European Laboratory for Particle Physics, CERN, 
     CH-1211 Geneva 23, Switzerland
\item[\wl] World Laboratory, FBLJA  Project, CH-1211 Geneva 23, Switzerland
\item[\geneva] University of Geneva, CH-1211 Geneva 4, Switzerland
\item[\hamburg] University of Hamburg, D-22761 Hamburg, Germany
\item[\hefei] Chinese University of Science and Technology, USTC,
      Hefei, Anhui 230 029, China$^{\triangle}$
\item[\lausanne] University of Lausanne, CH-1015 Lausanne, Switzerland
\item[\lyon] Institut de Physique Nucl\'eaire de Lyon, 
     IN2P3-CNRS,Universit\'e Claude Bernard, 
     F-69622 Villeurbanne, France
\item[\madrid] Centro de Investigaciones Energ{\'e}ticas, 
     Medioambientales y Tecnol\'ogicas, CIEMAT, E-28040 Madrid,
     Spain${\flat}$ 
\item[\florida] Florida Institute of Technology, Melbourne, FL 32901, USA
\item[\milan] INFN-Sezione di Milano, I-20133 Milan, Italy
\item[\moscow] Institute of Theoretical and Experimental Physics, ITEP, 
     Moscow, Russia
\item[\naples] INFN-Sezione di Napoli and University of Naples, 
     I-80125 Naples, Italy
\item[\cyprus] Department of Physics, University of Cyprus,
     Nicosia, Cyprus
\item[\nymegen] Radboud University and NIKHEF, 
     NL-6525 ED Nijmegen, The Netherlands
\item[\caltech] California Institute of Technology, Pasadena, CA 91125, USA
\item[\perugia] INFN-Sezione di Perugia and Universit\`a Degli 
     Studi di Perugia, I-06100 Perugia, Italy   
\item[\peters] Nuclear Physics Institute, St. Petersburg, Russia
\item[\cmu] Carnegie Mellon University, Pittsburgh, PA 15213, USA
\item[\potenza] INFN-Sezione di Napoli and University of Potenza, 
     I-85100 Potenza, Italy
\item[\prince] Princeton University, Princeton, NJ 08544, USA
\item[\riverside] University of Californa, Riverside, CA 92521, USA
\item[\rome] INFN-Sezione di Roma and University of Rome, ``La Sapienza",
     I-00185 Rome, Italy
\item[\salerno] University and INFN, Salerno, I-84100 Salerno, Italy
\item[\ucsd] University of California, San Diego, CA 92093, USA
\item[\sofia] Bulgarian Academy of Sciences, Central Lab.~of 
     Mechatronics and Instrumentation, BU-1113 Sofia, Bulgaria
\item[\korea]  The Center for High Energy Physics, 
     Kyungpook National University, 702-701 Taegu, Republic of Korea
\item[\taiwan] National Central University, Chung-Li, Taiwan, China
\item[\tsinghua] Department of Physics, National Tsing Hua University,
      Taiwan, China
\item[\purdue] Purdue University, West Lafayette, IN 47907, USA
\item[\psinst] Paul Scherrer Institut, PSI, CH-5232 Villigen, Switzerland
\item[\zeuthen] DESY, D-15738 Zeuthen, Germany
\item[\eth] Eidgen\"ossische Technische Hochschule, ETH Z\"urich,
     CH-8093 Z\"urich, Switzerland
\item[\S]  Supported by the German Bundesministerium 
        f\"ur Bildung, Wissenschaft, Forschung und Technologie.
\item[\ddag] Supported by the Hungarian OTKA fund under contract
numbers T019181, F023259 and T037350.
\item[\P] Also supported by the Hungarian OTKA fund under contract
  number T026178.
\item[$\flat$] Supported also by the Comisi\'on Interministerial de Ciencia y 
        Tecnolog{\'\i}a.
\item[$\sharp$] Also supported by CONICET and Universidad Nacional de La Plata,
        CC 67, 1900 La Plata, Argentina.
\item[$\triangle$] Supported by the National Natural Science
  Foundation of China.
\end{list}
}
\vfill


\newpage

%
%

\begin{table}
\begin{center}
\begin{tabular}{|c|c|c|c|c|c|}
\hline
 $\langle \sqrt{s}\rangle$ (Ge\kern -0.1em V)& \lum ($\pb$) & $N_D$ & $N_{MC}$ &
 $N_S$ & $N_B$ \\
\hline
188.6 &           156.4 &            11561 &           11559 &              11288 &          271 \\
191.6 & \phantom{0}29.7 &  \phantom{0}1976 & \phantom{0}1953 &    \phantom{0}1905 &\phantom{0}48 \\
195.6 & \phantom{0}83.7 &  \phantom{0}5677 & \phantom{0}5673 &    \phantom{0}5539 &          134 \\
199.5 & \phantom{0}83.5 &  \phantom{0}5382 & \phantom{0}5338 &    \phantom{0}5201 &          137 \\
201.8 & \phantom{0}39.1 &  \phantom{0}2379 & \phantom{0}2417 &    \phantom{0}2355 &\phantom{0}62 \\
205.2 & \phantom{0}75.9 &  \phantom{0}4259 & \phantom{0}4165 &    \phantom{0}4063 &          102 \\
206.7 &           130.4 &  \phantom{0}7388 & \phantom{0}7512 &    \phantom{0}7339 &          173 \\
208.2 & \phantom{00}8.7 &  \phantom{00}441 & \phantom{00}484 &    \phantom{00}473 &\phantom{0}11 \\
\hline
 198.0 & 607.4 & 39063 & 39101 & 38163 & 938 \\
\hline
\end{tabular}
\caption{\label{tab:1}Luminosity-averaged centre-of-mass energies, $\langle
  \sqrt{s}\rangle$, and corresponding integrated luminosities, \lum,
  used in the analysis. The $\rts$ spread in each point is of the
  order of $1\GeV$. The numbers of observed events, $N_D$, are given
  together with the total Monte Carlo expectations,
  $N_{MC}$, and their breakdown into signal, $N_S$, and background,
  $N_B$, events. The last row lists the average centre-of-mass energy,
  the total integrated luminosity and the total numbers of events.}
\end{center}
\end{table}

\begin{sidewaystable}
\begin{center}
\begin{tabular}{|c|c|c|c|c|c|c|c|c|c|}
\cline{3-10}
\multicolumn{1}{c}{}  &
\multicolumn{1}{c|}{}  &
\multicolumn{8}{c|}{${\rm d}\,\sigma/{\rm d}\, |\cos\theta|$ (pb)}  \\
\cline{3-10}
\multicolumn{1}{c}{}  &
\multicolumn{1}{c|}{}  &
\multicolumn{2}{c|}{$\langle\rts\rangle = 188.6 \GeV$} &
\multicolumn{2}{c|}{$\langle\rts\rangle = 191.6 \GeV$} &
\multicolumn{2}{c|}{$\langle\rts\rangle = 195.6 \GeV$} &
\multicolumn{2}{c|}{$\langle\rts\rangle = 199.5 \GeV$} \\
\hline
$|\cos\theta|$ & $\langle|\cos\theta|\rangle$ &
Meas. & Exp. &
Meas. & Exp. &
Meas. & Exp. &
Meas. & Exp. \\
\hline
$ 0.00- 0.09$ & 0.052 &
$ \phantom{0}12.8 \pm  \phantom{0}0.9 \pm \phantom{0}0.2$ & $ \phantom{0}10.4$ & 
$ \phantom{00}9.3 \pm  \phantom{0}1.9 \pm \phantom{0}0.2$ & $ \phantom{0}10.1$ & 
$ \phantom{00}8.6 \pm  \phantom{0}1.0 \pm \phantom{0}0.1$ & $ \phantom{00}9.6$ & 
$ \phantom{0}10.3 \pm  \phantom{0}1.1 \pm \phantom{0}0.2$ & $ \phantom{00}9.2$ \\
$ 0.09- 0.18$ & 0.138 &
$ \phantom{0}10.9 \pm  \phantom{0}0.8 \pm  \phantom{0}0.1$ & $ \phantom{0}11.3$ &
$ \phantom{0}10.2 \pm  \phantom{0}2.0 \pm  \phantom{0}0.2$ & $ \phantom{0}11.0$ &
$ \phantom{0}10.5 \pm  \phantom{0}1.2 \pm  \phantom{0}0.1$ & $ \phantom{0}10.5$ &
$ \phantom{00}9.8 \pm  \phantom{0}1.1 \pm  \phantom{0}0.1$ & $ \phantom{0}10.0$ \\
$ 0.18- 0.27$ & 0.227 &
$ \phantom{0}14.0 \pm  \phantom{0}1.0 \pm \phantom{0}0.2$ & $ \phantom{0}13.4$ &
$ \phantom{0}11.5 \pm  \phantom{0}2.1 \pm \phantom{0}0.2$ & $ \phantom{0}13.0$ &
$ \phantom{0}12.0 \pm  \phantom{0}1.2 \pm \phantom{0}0.2$ & $ \phantom{0}12.4$ &
$ \phantom{0}12.2 \pm  \phantom{0}1.3 \pm \phantom{0}0.2$ & $ \phantom{0}11.9$ \\
$ 0.27- 0.36$ & 0.317 &
$ \phantom{0}16.2 \pm  \phantom{0}1.0 \pm \phantom{0}0.1$ & $ \phantom{0}17.1$ &
$ \phantom{0}14.3 \pm  \phantom{0}2.4 \pm \phantom{0}0.2$ & $ \phantom{0}16.6$ &
$ \phantom{0}18.0 \pm  \phantom{0}1.5 \pm \phantom{0}0.2$ & $ \phantom{0}15.9$ &
$ \phantom{0}14.7 \pm  \phantom{0}1.4 \pm \phantom{0}0.2$ & $ \phantom{0}15.2$ \\
$ 0.36- 0.45$ & 0.407 &
$ \phantom{0}25.0 \pm  \phantom{0}1.3 \pm \phantom{0}0.2$ & $ \phantom{0}23.7$ &
$ \phantom{0}23.6 \pm  \phantom{0}3.0 \pm \phantom{0}0.3$ & $ \phantom{0}22.9$ &
$ \phantom{0}20.6 \pm  \phantom{0}1.6 \pm \phantom{0}0.2$ & $ \phantom{0}21.9$ &
$ \phantom{0}20.0 \pm  \phantom{0}1.6 \pm \phantom{0}0.2$ & $ \phantom{0}21.0$ \\
$ 0.45- 0.54$ & 0.497 &
$ \phantom{0}35.0 \pm  \phantom{0}1.6 \pm \phantom{0}0.3$ & $ \phantom{0}35.3$ &
$ \phantom{0}30.9 \pm  \phantom{0}3.5 \pm \phantom{0}0.3$ & $ \phantom{0}34.2$ &
$ \phantom{0}32.4 \pm  \phantom{0}2.1 \pm \phantom{0}0.3$ & $ \phantom{0}32.8$ &
$ \phantom{0}28.0 \pm  \phantom{0}1.9 \pm \phantom{0}0.3$ & $ \phantom{0}31.5$ \\
$ 0.54- 0.63$ & 0.588 &
$ \phantom{0}57.9 \pm  \phantom{0}2.0 \pm \phantom{0}1.1$ & $ \phantom{0}57.7$ &
$ \phantom{0}61.2 \pm  \phantom{0}5.0 \pm \phantom{0}1.2$ & $ \phantom{0}55.9$ &
$ \phantom{0}51.2 \pm  \phantom{0}2.6 \pm \phantom{0}1.0$ & $ \phantom{0}53.6$ &
$ \phantom{0}49.3 \pm  \phantom{0}2.6 \pm \phantom{0}1.0$ & $ \phantom{0}51.5$ \\
$ 0.63- 0.72$ & 0.678 &
$ 109.8 \pm \phantom{0}3.1 \pm  \phantom{0}2.6$ & $ 105.8$ &
$ 109.4 \pm \phantom{0}7.4 \pm  \phantom{0}2.9$ & $ 102.6$ &
$ \phantom{0}99.3 \pm   \phantom{0}4.0 \pm \phantom{0}2.5$ & $ \phantom{0}98.5$ &
$ \phantom{0}98.9 \pm   \phantom{0}4.1 \pm \phantom{0}2.5$ & $ \phantom{0}94.6$ \\
$ 0.72- 0.81$ & 0.770 &
$ 227.1 \pm   18.2 \pm   10.8$ & $ 232.2$ &
$ 196.4 \pm   39.3 \pm   16.3$ & $ 225.1$ &
$ 211.2 \pm   23.6 \pm   14.0$ & $ 216.2$ &
$ 231.3 \pm   26.9 \pm   16.2$ & $ 207.7$ \\
$ 0.81- 0.90$ & 0.862 &
$ 735.4 \pm  \phantom{0}8.4 \pm \phantom{0}6.3$ & $ 735.9$ &
$ 720.4 \pm   19.8 \pm \phantom{0}7.3$ & $ 713.5$ &
$ 690.4 \pm   11.2 \pm \phantom{0}6.4$ & $ 685.1$ &
$ 670.3 \pm   11.1 \pm \phantom{0}6.2$ & $ 658.4$ \\
\hline
\multicolumn{1}{c}{}  &
\multicolumn{1}{c|}{}  &
\multicolumn{2}{c|}{$\langle\rts\rangle = 201.8 \GeV$} &
\multicolumn{2}{c|}{$\langle\rts\rangle = 205.2 \GeV$} &
\multicolumn{2}{c|}{$\langle\rts\rangle = 206.7 \GeV$} &
\multicolumn{2}{c|}{$\langle\rts\rangle = 208.2 \GeV$} \\
\hline
$|\cos\theta|$ & $\langle|\cos\theta|\rangle$ &
Meas. & Exp. &
Meas. & Exp. &
Meas. & Exp. &
Meas. & Exp. \\
\hline
$ 0.00- 0.09$ &  0.052 &
$ \phantom{0}11.0 \pm \phantom{0}1.7 \pm \phantom{0}0.1$ & $ \phantom{00}9.0$ & 
$ \phantom{00}8.7 \pm \phantom{0}1.2 \pm \phantom{0}0.2$ & $ \phantom{00}8.8$ & 
$ \phantom{00}9.0 \pm \phantom{0}0.9 \pm \phantom{0}0.1$ & $ \phantom{00}8.6$ & 
$ \phantom{00}3.9 \pm \phantom{0}2.3 \pm \phantom{0}0.1$ & $ \phantom{00}8.5$ \\
$ 0.09- 0.18$ & 0.138 &
$ \phantom{0}11.0 \pm \phantom{0}1.7 \pm \phantom{0}0.1$ & $ \phantom{00}9.8$ &
$ \phantom{0}12.9 \pm \phantom{0}1.4 \pm \phantom{0}0.2$ & $ \phantom{00}9.6$ &
$ \phantom{00}8.7 \pm  \phantom{0}0.9 \pm \phantom{0}0.1$ & $ \phantom{00}9.4$ &
$ \phantom{0}10.3 \pm \phantom{0}3.7 \pm \phantom{0}0.2$ & $ \phantom{00}9.2$ \\
$ 0.18- 0.27$ & 0.227 &
$ \phantom{0}11.9 \pm \phantom{0}1.8 \pm \phantom{0}0.2$ & $\phantom{0}11.6$ &
$ \phantom{0}12.3 \pm \phantom{0}1.4 \pm \phantom{0}0.2$ & $ \phantom{0}11.4$ &
$ \phantom{0}10.4 \pm \phantom{0}0.9 \pm \phantom{0}0.2$ & $\phantom{0}11.2$ &
$ \phantom{00}5.4 \pm \phantom{0}2.7 \pm \phantom{0}0.1$ & $ \phantom{0}10.9$ \\
$ 0.27- 0.36$ & 0.317 &
$ \phantom{0}14.8 \pm \phantom{0}2.1 \pm \phantom{0}0.2$ & $ \phantom{0}14.9$ &
$ \phantom{0}16.1 \pm \phantom{0}1.6 \pm \phantom{0}0.1$ & $ \phantom{0}14.6$ &
$ \phantom{0}16.8 \pm \phantom{0}1.2 \pm \phantom{0}0.2$ & $ \phantom{0}14.3$ &
$ \phantom{0}13.4 \pm \phantom{0}4.2 \pm \phantom{0}0.1$ & $ \phantom{0}14.0$ \\
$ 0.36- 0.45$ &  0.407 &
$ \phantom{0}21.2 \pm \phantom{0}2.5 \pm \phantom{0}0.2$ & $ \phantom{0}20.6$ &
$ \phantom{0}20.0 \pm \phantom{0}1.8 \pm \phantom{0}0.2$ & $ \phantom{0}20.2$ &
$ \phantom{0}23.1 \pm \phantom{0}1.4 \pm \phantom{0}0.3$ & $ \phantom{0}19.8$ &
$ \phantom{0}23.4 \pm \phantom{0}5.7 \pm \phantom{0}0.3$ & $ \phantom{0}19.4$ \\
$ 0.45- 0.54$ &  0.497 &
$ \phantom{0}37.2 \pm \phantom{0}3.3 \pm \phantom{0}0.4$ & $ \phantom{0}30.9$ &
$ \phantom{0}31.7 \pm \phantom{0}2.3 \pm \phantom{0}0.4$ & $ \phantom{0}30.2$ &
$ \phantom{0}29.4 \pm \phantom{0}1.6 \pm \phantom{0}0.4$ & $ \phantom{0}29.7$ &
$ \phantom{0}26.3 \pm \phantom{0}6.0 \pm \phantom{0}0.3$ & $ \phantom{0}29.1$ \\
$ 0.54- 0.63$ &  0.588 &
$ \phantom{0}55.5 \pm \phantom{0}4.1 \pm \phantom{0}1.0$ & $ \phantom{0}50.5$ &
$ \phantom{0}48.0 \pm \phantom{0}2.8 \pm \phantom{0}0.9$ & $ \phantom{0}49.5$ &
$ \phantom{0}44.5 \pm \phantom{0}2.0 \pm \phantom{0}0.8$ & $ \phantom{0}48.5$ &
$ \phantom{0}37.6 \pm \phantom{0}7.2 \pm \phantom{0}0.7$ & $ \phantom{0}47.6$ \\
$ 0.63- 0.72$ & 0.678 &
$ \phantom{0}91.4 \pm \phantom{0}5.7 \pm \phantom{0}2.3$ & $ \phantom{0}92.7$ &
$ \phantom{0}93.3 \pm \phantom{0}4.3 \pm \phantom{0}2.3$ & $ \phantom{0}90.9$ &
$ \phantom{0}90.0 \pm \phantom{0}3.2 \pm \phantom{0}2.4$ & $ \phantom{0}89.2$ &
$ \phantom{0}84.3 \pm   12.0 \pm   \phantom{0}2.3$ & $ \phantom{0}87.5$ \\
$ 0.72- 0.81$ &0.770 &
$ 243.7 \pm   39.0 \pm   18.7$ & $ 203.7$ &
$ 252.2 \pm   29.3 \pm   15.5$ & $ 199.7$ &
$ 170.0 \pm   17.5 \pm   14.5$ & $ 195.9$ &
$ 280.3 \pm   88.6 \pm   24.0$ & $ 192.2$ \\
$ 0.81- 0.90$ & 0.862 &
$ 618.3 \pm   15.8 \pm  \phantom{0}6.0$ & $ 645.7$ &
$ 628.9 \pm   11.9 \pm  \phantom{0}5.7$ & $ 633.3$ &
$ 604.4 \pm  \phantom{0}8.7 \pm \phantom{0}6.2$ & $ 621.2$ &
$ 565.3 \pm   33.0 \pm   \phantom{0}5.8$ & $ 609.5$ \\
\hline
\end{tabular}
\end{center}
\caption{\label{tab:2}Measured, Meas., and expected, Exp., folded
  differential cross sections for the eight average centre-of-mass
  energies, $\langle \sqrt{s}\rangle$, and the ten $|\cos\theta|$
  intervals, with expected average values
  $\langle|\cos\theta|\rangle$. The first uncertainty is statistical
  and the second systematic.  }
\end{sidewaystable}

\begin{table}
\begin{center}
\begin{tabular}{|c|c|c|c|}
\hline
$\langle|\cos\theta|\rangle$ & ${\rm d}\,\sigma \over{\rm d}\,
  |\cos\theta|$ (pb)& ${\rm d}\,\sigma^{th}(1) \over{\rm d}\,
  |\cos\theta|$ (pb)& ${\rm d}\,\sigma^{th}(0) \over {\rm d}\,
  |\cos\theta|$ (pb)\\
\hline
 0.052 & $\phantom{00}9.93	        \pm 0.42 \pm	        \phantom{0}0.15	        $ &  $\phantom{00}9.7 $&  $ \phantom{00}8.6 $ \\
 0.138 & $\phantom{0}10.25	        \pm 0.43 \pm	        \phantom{0}0.21	        $ &  $\phantom{0}10.5 $&  $ \phantom{00}9.4 $ \\
 0.227 & $\phantom{0}11.99	        \pm 0.47 \pm	        \phantom{0}0.14            $ &  $\phantom{0}12.4 $&  $ \phantom{0}11.0 $ \\
 0.317 & $\phantom{0}15.95	        \pm 0.54 \pm	        \phantom{0}0.14	        $ &  $\phantom{0}15.8 $&  $ \phantom{0}14.2 $ \\
 0.407 & $\phantom{0}22.15	        \pm 0.64 \pm            \phantom{0}0.25	        $ &  $\phantom{0}21.7 $&  $ \phantom{0}19.7 $ \\
 0.497 & $\phantom{0}31.65	        \pm 0.77 \pm	        \phantom{0}0.17	        $ &  $\phantom{0}32.2 $&  $ \phantom{0}29.5 $ \\
 0.588 & $\phantom{0}51.15   	        \pm 0.99 \pm	        \phantom{0}0.26            $ &  $\phantom{0}52.3 $&  $ \phantom{0}48.4 $ \\
 0.678 & $\phantom{0}98.7\phantom{0}	\pm 1.5\phantom{0} \pm	\phantom{0}1.2\phantom{0}  $ &  $\phantom{0}95.8 $&  $ \phantom{0}89.1 $ \\
 0.770 & $211.6\phantom{0}	        \pm 9.1\phantom{0} \pm	13.9\phantom{0} $ &  $     210.2      $&  $	197.0	    $ \\
 0.862 & $666.9\phantom{0} 	        \pm 4.1\phantom{0} \pm	\phantom{0}4.9\phantom{0}  $ &  $     671.1      $&  $	634.2       $ \\
\hline
\end{tabular}
\caption{\label{tab:3}Combined differential cross sections for the
  luminosity-averaged centre-of-mass energy $\langle \sqrt{s}\rangle=198\GeV$, compared with the
  Standard Model expectations, ${\rm d}\,\sigma^{th}(1)/{\rm d}\,
  |\cos\theta|$, and the expectations for the case
  in which $\alpha$ does not change with $Q^2$, ${\rm d}\,\sigma^{th}(0)/{\rm d}\,
  |\cos\theta|$. The first uncertainties are
  statistical and
  the second systematic.}
\end{center}
\end{table}

\begin{table}
\begin{center}
\begin{tabular}{|l|c|}
\hline 
Source of uncertainty & $\Delta C$\\
\hline
Theoretical uncertainty  & \,\,\,\,\,\,\,0.11 \\
Experimental systematic  & \,\,\,\,\,\,\,0.08 \\
$F_{rad}$                & \,\,\,\,\,\,\,0.05 \\
Bin migration            & $<0.01$ \\
\hline
Total & \,\,\,\,\,\,\,0.14 \\
\hline
\end{tabular}
\caption{\label{tab:4}
Sources of systematic uncertainty and their effect, $\Delta C$, on the
determination of the $C$ parameter.}
\end{center}
\end{table}

\clearpage

%
%

\end{document}